\documentclass{PoS}
\usepackage{wrapfig}
\usepackage{amssymb}

\title{ Global analysis on determination of fracture functions considering sea quark asymmetries in the nucleon }

\ShortTitle{ Global analysis on determination of fracture functions }

\author{\speaker{Samira Shoeibi} \\
Department of Physics, Ferdowsi University of Mashhad, P.O.Box 1436, Mashhad, Iran  \\
E-mail: \email{Samira.Shoeibimohsenabadi@mail.um.ac.ir}}

\author{F. Taghavi-Shahri \\
Department of Physics, Ferdowsi University of Mashhad, P.O.Box 1436, Mashhad, Iran  \\
E-mail: \email{Taghavishahri@um.ac.ir}}

\author{Kurosh Javidan \\
Department of Physics, Ferdowsi University of Mashhad, P.O.Box 1436, Mashhad, Iran  \\
E-mail: \email{Javidan@um.ac.ir}}

\abstract
{
Several experiments at the electron-proton ($ep$) collider HERA have collected high precision data on the spectrum of leading-proton and leading-neutron carrying a large fraction of the proton's energy. In this paper, we have analyzed recent experimental data on the production of leading-nucleon in deep inelastic scattering (DIS) processes at HERA in the framework of a perturbative QCD (pQCD). An approach based on the fractures functions framework has been used, and the nucleon fracture functions (nucleon FFs) ${\cal M}_2^{(n/p)} (x, Q^2, x_L)$ have been extracted from global QCD analysis of DIS data measured by ZEUS collaboration at HERA. We show that the approach of fracture functions formalisem allows us phenomenologically parametrize the nucleon FFs at the input scale, $Q_0^2$.
Considering leading-nucleon production data in the DIS processes, we present the results for the separate parton distributions for all parton species. 
The extracted results from the $t$-integrated leading-baryon fracture functions, $F_2^{\rm LB(3)} (x, Q^2, x_L)$ are in good agreement with all DIS data analyzed, for a wide range of longitudinal momentum fraction $x_L$ as well as scaled fractional momentum variable $x$.
}

\FullConference{ XVII INTERNATIONAL CONFERENCE ON HADRON SPECTROSCOPY AND STRUCTURE 	\\
			September 25th-29th 2017 \\
		Salamanca, Spain  }

\begin{document}

\section{Introduction}

In recent years, due to considerable attention toward semi-inclusive cross sections to the leading-baryon productions at HERA, different phenomenological groups try to propose standard parametric form for the parton densities of the incoming hadrons to describe the leading-baryon production processes. Leading-baryons carry a sizable fraction of the incoming hadron energy and they are produced at small polar angle with respect to the collision axis, ($\theta_{B}=0.8$ mrad), in the target fragmentation region.
In our recent analysis~\cite{Shoeibi:2017lrl}, we have used leading-neutron production data measured by the H1~\cite{Aaron:2010ab} and ZEUS~\cite{Chekanov:2002pf} collaborations to study the fracture functions formalism which allows us to phenomenologically parametrize the neutron FFs at the input scale, $Q_0^2$. We have studied the determination of singlet and gluon distributions of neutron FFs, with no assumption on the separation between quark and anti-quark FFs. However, In the present analysis, in order to consider fracture functions formalism more precisely, we study the asymmetry of quark and antiquark by using ZEUS collaboration data on leading-neutron and leading-proton production~\cite{Chekanov:2002pf,Rinaldi:2006mf,Chekanov:2008tn}.

\section{Fracture Functions}

In order to describe those hadrons produced in the target fragmentation region~~\cite{Trentadue:1993ka,Trentadue:1994iw}, new distribution functions should be introduced.
This new quantity is an un-calculable, but measurable and universal functions, the so-called Fracture Functions (FFs), ${\cal M}_{i}^{\frac{h}{N}}(x, z, Q^{2})$~\cite{Trentadue:1993ka,Trentadue:1994iw}. 
This functions represent the conditional probability of finding a parton $i$ (with momentum fraction $x$ of incoming hadron), while a hadron $h$ with momentum fraction $z$ is detected. The scale dependence of fracture functions at $O(\alpha_{s})$ in fixed value of $z$ obey standard DGLAP evolutions equations~\cite{deFlorian:1997wi}:

\begin{eqnarray}\label{Eq:dglap1}
\frac{\partial}{\partial logQ^2} \, {\cal M}_{i, h/N}(x, z, Q^2) = \frac{\alpha_{s}(Q^2)}{2 \pi} \int^1_{\frac{x}{1-z}} \frac{du}{u} P^i_j(u) \, {\cal M}_{j, h/N}(\frac{x}{u},z ,Q^2) + \nonumber \\
\frac{\alpha_{s}(Q^2)} {2 \pi} \int^{\frac{x}{x + z}}_{x}\frac{du}{(1 - u)}P^{i,l}_j(u) \, f_{j/p}(\frac{x}{u}, Q^2) \, D_{h/l} (\frac{z u}{x (1 - u)}, Q^2) 
\end{eqnarray}

where $f_{j/p}$ is the parton distribution of flavor $j$, $D_{h/l}$ is the fragmentation function of hadron $h$ from a parton $l$, $P^i_j(u)$  and $P^{(i,l)}_j$ are the regularized~\cite{Altarelli:1977zs} and real~\cite{Konishi:1979cb} Altarelli-Parisi (AP) splitting functions, respectively.

\section{Forward baryons productions at HERA}

As we mentioned in the Introduction, ZEUS Collaboration has measured DIS events in which neutron and proton are produced in the forward region~\cite{Chekanov:2002pf,Rinaldi:2006mf,Chekanov:2008tn}. In addition to kinematical variables used to described DIS processes ($Q^{2}=-q^{2}, \, x=\frac{Q^{2}}{2p.q}$ and $y=\frac{p.q}{p.k}$), two more kinematical variables also needed to describe the final state nucleon. These are the longitudinal momentum fraction $x_{L}$ and the squared four-momentum transfer $t$ between the incident proton and the final state nucleon, which are given by

\begin{eqnarray}
x_{L}  =  1 - \frac{q.(p - p_{n})}{q.p} \simeq \frac{E_{B}} {E_{P}}\,, \, \, \, \, \, \, {\rm and} \, \, \, \, \, \,  t  =  (p - p_{B})^{2}\simeq -\frac{p_{T}^{2}}{x_{L}} - (1 - x_{L}) (\frac{m_B^2}{x_{L}} - m_{p}^{2}) \,,  \nonumber
\end{eqnarray}

where $m_p$ is the proton mass, $p_B$ is the four-momentum of the final state baryons, $m_B$ is the nucleon mass, and $E_B$ and $p_T$ are the nucleon energy and transverse momentum, respectively. The four-fold differential cross section for leading-Baryon production, which can be written as semi-inclusive structure function, $F_2^{LB(4)}$, is given by:

\begin{eqnarray}
\label{eq:forward1}
\frac{d^{4} \sigma (e p \rightarrow e B X)} {dx dQ^{2} dx_L dp^{2}_{T}} = \frac{4 \pi \alpha^{2}}{x Q^{2}} (1 - y + \frac{y^2}{2}) \, F_2^{LB(4)}(x, Q^{2}, x_{L}, p^{2}_{T}) \,.
\end{eqnarray}

where $F_2^{LB(4)} (x, Q^{2}, x_{L}, p^{2}_{T})$ is leading-baryon transverse structure functions.

\section{Leading-nucleon structure functions}

The leading-nucleon transverse structure functions can be written in terms of "fracture functions" and hard-scattering coefficient functions~\cite{Shoeibi:2017lrl,Shoeibi:2017zha,Ceccopieri:2014rpa} as,

\begin{eqnarray}
\label{eq:factorization}
F^{LB(3)} (x, Q^2; x_L, p_T^2)  = \sum_{i} \int_{x}^{1} \frac{d \xi}{\xi} {\cal M}^B_{i/p} (x, \mu_F^2; x_L, p_T^2)  \times C_i (\frac{x}{\xi}, \frac{Q^2}{\mu_F^2}, \alpha_s(\mu^2_R)) + {\cal O}  (\frac{1}{Q^2}) \,.
\end{eqnarray}

The index $i$ runs on the flavour of the interacting parton and the Wilson coefficient functions, $C_q$ and $C_g$, are the same as in fully inclusive DIS~\cite{Vermaseren:2005qc}. By integrating over $dp^2_T$ up to $p_T^{\rm max}$, gives

\begin{equation}
\label{intpT}
F_2^{LB(3)}(x, x_L, Q^2)  = \int_0^{p^{2 \, max}_T} F_2^{LB(4)} (x, x_L, Q^2, p^2_T)  dp^2_T \,.
\end{equation}

The resulting $p_{T}$-integrated leading-nucleon fracture function

\begin{equation}
\label{fracint}
{\cal M}_{i/p}^{B}(x, Q^{2}, x_{L}) = \int^{p^{2}_{T, max}}dp_{T}^{2} {\cal M} _ {i/p}^{B}(x, Q^{2}, x_{L}, p_{T}^{2})\,.
\end{equation}

obeys the well-known DGLAP evolution equations

\begin{equation}
\label{eq:dglap2}
Q^{2} \frac{\partial {\cal M}_{\frac{i}{P}}^{B}(x,Q^{2},x_{L})}{\partial Q^{2}}=\frac{\alpha_{s}(Q^{2})}{2\pi}\int_{x}^{1}\frac{du}{u}P^{j}_{i}(u) {\cal M}_\frac{j}{P}^{B}(\frac{x}{u},Q^{2},x_{L})
\end{equation}

\section{ Hypothesis of limiting fragmentation }

The ``hypothesis of limiting fragmentation''~\cite{Benecke:1969sh,Chou:1994dh}, implies that final state baryons is not depend to the $x$ and $Q^2$ and it also states that target fragmentation is independent of the incident projectile's energy. Therefore, using this hypothesis we can write the leading-nucleon transverse structure functions as:

\begin{equation}
\label{fracint}
F_{2}^{LB(4)}(x, Q^{2}, x_{L}, p_{T})=f(x_{L}, p_{T}) \times F(x, Q^{2})\,.
\end{equation}

This hypothesis has been tested by H1 Collaboration for leading-proton and leading-neutron data~\cite{Adloff:1998yg} and it is also used to extract pion structure function from leading neutron electroproduction~\cite{McKenney:2015xis}.

\section{ Parameterization of nucleon FFs }

Note that our goal is to present nucleon FFs from QCD analysis of leading-baryon production in the semi-inclusive DIS reaction $e p \rightarrow e B X$ at HERA.
We choose the standard functional form for the nucleon FFs at the input scale $Q_0^2 = 2 \, {\rm GeV}^2$ ,

\begin{eqnarray}
\label{eq:PDFQ0-New}
x{\cal M}^{n}_{f}(x ,Q_{0}^{2}, x_{L}) &=& {\cal W}_{f}(x_{L}) \, xf^{\rm GJR08}(x,Q^{2})  \,, \nonumber \\
x{\cal M}^{n}_{g/P}(x, Q_{0}^{2}, x_{L}) &=& {\cal W}_g(x_{L}) \,  xg^{\rm GJR08}(x,Q^{2})   \,
\end{eqnarray}
where $xf \equiv xu_v, \, xd_v, \, x\Delta, \, xs=x\overline{s}, \, x(\bar{d}+\bar{u})$ and $ xg$ are the {\tt GJR08} parton distributions functions (PDFs)~\cite{Gluck:2007ck}. The weight functions ${\cal W}_f(x_L)$ and ${\cal W}_g(x_L)$ are given by

\begin{eqnarray}
\label{eq:W-New}
&&	{\cal W}_{f}(x_L) = {\cal N}_{f} \,\, x_L^{A_{f}} (1 - x_L)^{B_{f}} ( 1 + C_{f} \, x_L^{D_{f}} )  \,, \nonumber \\
&&	{\cal W}_g(x_L) = {\cal N}_g \,\, x_L^{A_g} (1 - x_L)^{B_g} ( 1 + C_g \, x_L^{D_g} )  \,, \nonumber
\end{eqnarray}

where $p_i = \{{\cal N}_i, A_i, B_i, C_i, D_i\}$ are the free parameters to be fitted. Experimental data sets that are used in our global analysis are shown in Table~\ref{tab:tabledata}.  The number of data points, and fitted normalization shifts ${\cal{N}}_{n}$ obtained in the fit are also presented as well.

\begin{table*}[htb]
\caption{List of all the leading-neutron and leading-proton productions data points used in our global analysis. The number of data points, and fitted normalization shifts ${\cal{N}}_{n}$ obtained in the fit are also presented as well.} \label{tab:tabledata}
\begin{tabular}{l c c c c c c}
Experiment & Observable & [$x_L^{\rm min}, x_L^{\rm max}$]  & $Q^2\,[{\rm GeV}^2]$  & \# of points & ${\cal N}_n$
\tabularnewline
\hline\hline
ZEUS-02~\cite{Chekanov:2002pf} & $F_2^{LN(3)}$ &   [0.24--0.92]    & 7--1000  & \textbf{300} &  0.9974   \\		
ZEUS-06~\cite{Rinaldi:2006mf}  & $r^{LP(3)}$   &   [0.575--0.890]  & 3.4--377 & \textbf{226} &  1.0012  \\		
ZEUS-09~\cite{Chekanov:2008tn} & $r^{LP(3)}$   &   [0.370--0.895]  & 4.2--237 & \textbf{168} &  1.0004  \\		
\hline \hline
\multicolumn{1}{c}{\textbf{Total data points}}  &  &  &  &  \textbf{694}  \\  \hline
\end{tabular}
\end{table*}

Using {\tt QCDNUM17} package~\cite{Botje:2010ay}, the distribution functions in Eq.~\ref{eq:PDFQ0-New} are evolved within a Zero-Mass Variable Flavour Number Scheme (ZM-VFNS) at next-to-leading order (NLO). The minimization has been done using the CERN {\tt MINUIT} package~\cite{James:1975dr}. In our analysis, we minimized the $\chi^2_{\rm global}(\{\xi_i\})$ function with the free unknown parameters. This function is given by~\cite{Shoeibi:2017lrl,Shoeibi:2017zha,Shahri:2016uzl}

\begin{equation}\label{eq:chi}
\chi_{\rm global}^{2} (\{\xi_{i}\}) = \sum_{n = 1}^{n^{\rm exp}} w_{n} \chi_{n}^{2}\,,
\end{equation}

where $w_n$ is a weight factor for the $n^{\rm th}$ experiment and

\begin{eqnarray}\label{eq:chiglobal}
\chi_n^2 (\{\xi_{i}\}) = \left( \frac{1 -{\cal N}_{n} } {\Delta{\cal N}_{n}}\right)^2 + \sum_{j = 1}^{N_{n}^{\rm data}} \left(\frac{ ( {\cal N}_{n}  \, {D}_j^{\rm data} - {T}_j^{\rm theory}(\{\xi_{i}\})}{{\cal N}_{n} \, \delta {D}_{j}^{data}} \right)^2\,,
\end{eqnarray}

where $n^{exp}$ is correspond to the individual experimental data sets, and $N^{data}_n$ corresponds to the number of data points in each data set. The uncertainties of nucleon FFs as well as the leading-nucleon structure functions have been obtained using the Hessian method~\cite{Shahri:2016uzl,Khanpour:2016pph,Khanpour:2016uxh,Martin:2009iq,MoosaviNejad:2016ebo,Khanpour:2017cha,Accardi:2016ndt,Khanpour:2017fey,Soleymaninia:2017xhc}.

\section{Results}

\begin{figure}[h!]
\centerline{\includegraphics[width=16cm]{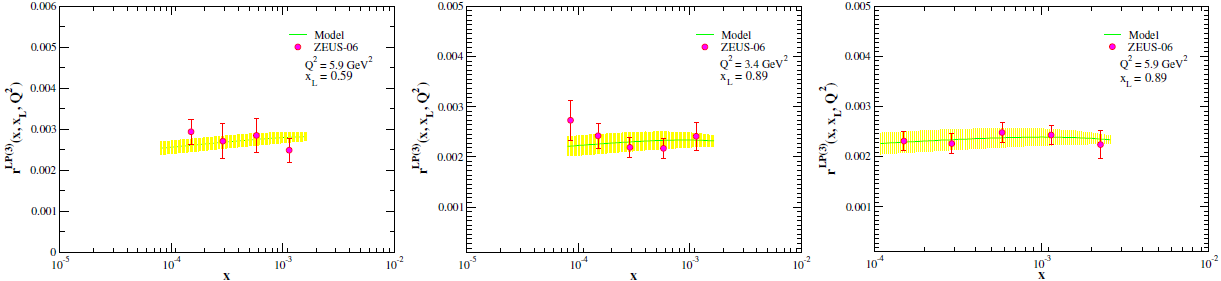}}
\caption{ (Color online) Our theory predictions for the structure function ratio $r^{LP(3)} (x, x_{L}, Q^{2})  = \frac{F_2^{LP(3)}(x, x_{L}, Q^{2})}{F_2^p(x, Q^{2})}$ and their uncertainties at 68\% C.L. in comparison with the ZEUS-06 data~\cite{Rinaldi:2006mf}. }\label{fig:Ratio}
\end{figure}

\begin{figure}[h!]
\centerline{\includegraphics[width=16cm]{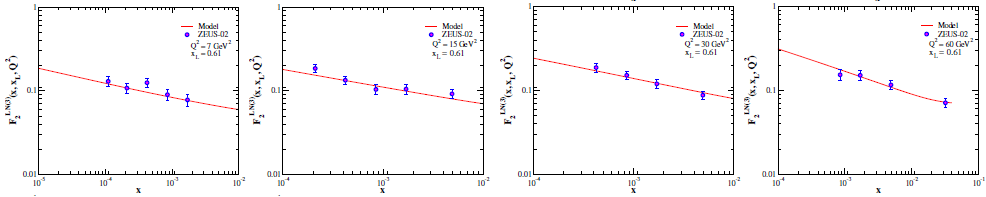}}
\caption{ (Color online) The tagged-neutron structure function $F^{LN(3)}_{2} (x, x_{L}, Q^{2})$ as a function of $x$ for some selected values of Q$^2$ at fixed value of $x_L = 0.61$. Our theory predictions have been compared with the ZEUS-02 leading neutron data~\cite{Chekanov:2002pf}. }\label{fig:Comparison2}
\end{figure}

Parameter values $\{\xi_i\}$ for our QCD analysis at the input scale $Q_0^2 = 2  \, {\rm GeV}^2$ obtained from QCD fit and nucleon FFs $x {\cal M}_{i} (x, x_{L}, Q^{2})$ for all parton species resulting from our QCD analysis are presented in Ref.~\cite{Shoeibi:2017zha}. In Figs.~\ref{fig:Ratio} and \ref{fig:Comparison2}, we compare our theory predictions with the analyzed leading-nucleon datasets, consistent results are found.


\section{Summary}

In recent years, several experiments at the electron-proton collider HERA have collected high-precision data on the spectrum of leading-nucleon carrying a large fraction of the proton's energy. In addition to the experimental efforts, much successful phenomenology has been developed in order to understand the mechanism of leading-nucleon productions. We have presented our NLO QCD analysis of nucleon FFs using available data on the forward nucleon production at HERA. It is shown that our approach based on the fracture functions formalism allows us to phenomenologically parametrize the nucleon FFs at a given input scale, $Q_0^2$.

\section*{Acknowledgments}

Authors are especially grateful Hamzeh Khanpour for many useful discussions and comments. Fatemeh Taghavi-Shahri and Kurosh Javidan also acknowledge Ferdowsi University of Mashhad.


\end{document}